# Infocast: A New Paradigm for Collaborative Content Distribution from Roadside Units to Vehicular Networks

Mohsen Sardari[†], Faramarz Hendessi[‡], and Faramarz Fekri[†]

[†]School of Electrical and Computer Engineering, Georgia Institute of Technology, Atlanta, GA 30332

[‡]Department of Electrical and Computer Engineering, Isfahan University of Technology, Isfahan, Iran

Email: {mohsen.sardari, fekri}@ece.gatech.edu, hendessi@cc.iut.ac.ir

*Abstract*—In this paper, we address the problem of distributing a large amount of bulk data to a sparse vehicular network from roadside infostations, using efficient vehicle-to-vehicle collaboration. Due to the highly dynamic nature of the underlying vehicular network topology, we depart from architectures requiring centralized coordination, reliable MAC scheduling, or global network state knowledge, and instead adopt a distributed paradigm with simple protocols. In other words, we investigate the problem of reliable dissemination from multiple sources when each node in the network shares a limited amount of its resources for cooperating with others. By using *rateless* coding at the Road Side Unit (RSU) and using vehicles as data carriers, we describe an efficient way to achieve reliable dissemination to all nodes (even disconnected clusters in the network). In the nutshell, we explore vehicles as mobile storage devices. We then develop a method to keep the density of the rateless codes packets as a function of distance from the RSU at the desired level set for the target decoding distance. We investigate various tradeoffs involving buffer size, maximum capacity, and the mobility parameter of the vehicles.[1]

## I. INTRODUCTION

Vehicular ad hoc networks (VANETs) have recently received considerable attention. Several applications developed for VANETs rely on data dissemination from an information source to many vehicles on the road. Although disseminating data from a server to a large number of clients has been studied in the database and the network community [1], [2], many unique characteristics of the VANET motivate to revisit some of the ideas. Vehicular communications have many different facets. Applications range from safety support [3], to entertainment for passengers, to local news delivery and advertisement [4]. The data traffic that drives VANETs ranges from short but critical public safety data with very tight latency constraints, to large amounts of bulk "download" data with relatively lenient latency constraints, to moderate amounts of near real-time data with somewhat tight latency constraints (e.g., streaming applications). It is not possible to address all aspects of vehicular communications with a single architectural proposal, and there is still need for fundamental understanding of the key properties of vehicular communication systems. In this work, by considering a simple but basic case we take some steps to answer questions concerning data dissemination in general vehicular networks such as the impact of mobility on throughput and reliability, robustness, and latency supported by such networks. Specifically, we will address those questions in the context of disseminating information packets from a large array of Roadside Units (RSU) to a bidirectional linear highway vehicular network. We assume each RSU is an independent source having a block of information packets to disseminate to all vehicles. Ideally, every vehicle must recover information data belonging to each RSU at any distance. However, this is not feasible given the resources. Hence, we are particularly interested in optimizing the distance from the source (RSU) where a typical vehicle can recover its data. This distance depends on the maximum throughput that can be achieved using efficient collaborative store-carry-forward routing. The application motivated this problem is the sale advertisement by the stores in an urban area.

Although disseminating data has been visited in the past in the network community, many unique characteristics of the VANET open up new research challenges. First, VANETs can be considered as a category of partitioned ad hoc networks [5]–[8]. Since density of vehicles is highly variable with space and time, the network changes from a sparsely disconnected network to a densely connected one in a short period of time. As a result of these topology variations, traditional routing and forwarding methods do not perform well in VANETs. Furthermore, many structures for efficient data dissemination such as trees, clustering, and grids, are extremely hard to set up and maintain in VANETs. In this paper we present a new approach that merges the *vehicle-to-vehicle* and *roadside-to-vehicle* communication typologies in order to support reliable data dissemination without the need of complex routing protocols. We suggest the application of a new class of packet-level coding schemes referred as *rateless* codes (Section III) for the reliable and efficient data dissemination in VANETs. Several aspects of rateless codes make them suitable for such applications. First, their rateless property avoids issues regarding the choice of rates even in the presence of varying link loss conditions. Second, rateless codes require very low coding overhead to recover the message while having low encoding and decoding complexity [9]. Using simulations, we show that the gains of rateless coding over classical store-and-forward multihop routing strategies is significant, as measured by the number of packets received at a vehicle as a function of the distance between the vehicle and the nearest RSU. We discuss as to how the limited resources (e.g. buffer size) available at each node in the network must be utilized for cooperation efficiently. Physical layer details, specific protocol formats, and all other details that are not relevant to the fundamental properties are coarsely modeled, while mobility will be addressed in more detail.

[1]This material is based upon work supported by the National Science Foundation under Grant CCF-0728772.

The rest of the paper is organized as follows. In Section III we provide an introduction to rateless codes. Section IV presents network assumptions, models and the details of our proposed scheme. Simulation results and discussion follow in Section V. We present future directions and conclude our paper in Section VI.

## II. RELATED WORK

The problem of distributing data from an infostation to vehicles on a highway is examined in [10]. The paper focuses on a simplified scenario where mobile nodes adopt a slotted ALOHA MAC strategy to cooperatively distribute content provided by a single infostation. The proposed architecture is designed for a dense network and will fail under realistic vehicle mobility. In [11], the performance of network coding as a solution for video streaming in vehicular networks is investigated. The paper adopts the 802.11 DCF MAC with network coding and studies, by means of computer simulations, its packet delivery ratio compared to the on demand multicast routing protocol (ODMRP) [12].

Based on the structure of the network and the amount of data generated in the network, broadcasting easily leads to severe congestion and significantly reduces the data delivery ratio. To solve such problems, Zhao et al. [13] proposed Data Pouring (DP) and buffering on the road scheme. The DP scheme explores the partially predictable vehicle mobility limited by the road layout. In DP, the data center broadcasts the data to one or more roads from the main roads going through it. Further, the vehicles on that road keep the broadcast data while they are on the road. The main road acts like the virtual buffer of the data center. Thus, vehicles in other crossing roads (when they are about to cross the main road) have the opportunity to collect data from the data center on the main street. In this scheme it was assumed that the vehicle density on the road is large enough to maintain network connectivity.

It should be noted that all these results were obtained in a scenario characterized by single source and infinite storage buffers. The problem of data dissemination involving multiple sources with finite resources has not been investigated. In this work we include the effect of limited resources for cooperation by considering a limited cooperation-buffer at each node in the network. In our "local diversity" recovery strategy, nodes only use this buffer for the data dissemination. We note that although the vehicles may have large storage buffers available, they may only use a small portion for helping any particular RSU. Our work is motivated by some of the questions rise here:

1) Is there any scheme that can handle both dense and sparse scenarios?
2) What is the optimal strategy for deleting packets from the buffers to satisfy the spatial relevance of the information?
3) What is the optimal method to use the buffers to handle a multisource dissemination scenario?
4) What are broadcast throughput and reliability and how one can improve them?

We try to take initial steps towards answering such questions.

## III. OVERVIEW OF RATELESS CODES

Also known as fountain codes, rateless codes were first conceived by Michael Luby [9]. As the name suggests, these codes are unlike conventional codes generated from algebraic and combinatorial means in the sense that they do not possess any fixed rate. From a finite set of data packets, the encoder can potentially generate an infinite stream of encoded packets. In [9], it was also shown that for $k$ data (information) packets, on the average, the destination requires $k\Gamma_k$ encoded packets, where $\Gamma_k = 1 + O(k^{-1/2} \log \frac{k}{\delta})$ is the overhead, to decode all the $k$ data packets with a probability of $1 - \delta$. Moreover, the encoding and decoding processes introduce very low computational complexity and are performed in the following manner. A parameter that is key in the design of rateless codes is the degree distribution polynomial $\Omega(x) = \sum_{1 \le x \le k} \Omega(i) x_i$ where $\Omega(i) \in [0, 1]$ for $i = 1, \ldots, k$. This degree distribution induces a probability distribution on the set of data packets $\{p_1, \ldots, p_k\}$ in the following manner. For any subset $V$ of packets, $P_\Omega(V) = \frac{\Omega(|V|)}{\binom{k}{|V|}}$. To generate a packet, the encoder generates an instance of a random variable $Z$ that selects each subset $V$ of packets with the aforementioned probability. Such a selection can be effected by equivalently selecting the weight $|V|$ of the selection using the distribution $\Omega$ and then selecting $|V|$ packets uniformly at random from set of $k$ data packets. To generate the encoded packet, the encoder does a packet-level XOR of the selected packets and appends each packet with the indices (or IDs) of all the packets XORed to generate the encoded packet. Figure 1 illustrates the encoding process. In this example, the encoded packets $e_1, e_2$ and $e_3$, are formed by selecting 3, 4 and 2 packets, respectively, from the set of $k$ information packets. In each encoded packet $e_i$, the dark portion (called the packet overhead) contains information of the indices of the packets used for generating $e_i$.

To decode the data packets from the received packets $e_i$, the decoder employs iterative message passing algorithm. To decode $k$ data packets, at least $k$ encoded packets must be collected at the receiver. However, in practical coding schemes with small $k$ (of the order of $10^3$), more than $k$ packets are needed for successful decoding with high probability. The ratio of the required number of distinct encoded packets to the number of data packets $k$ is referred as the *coding overhead* $\Gamma_k$.

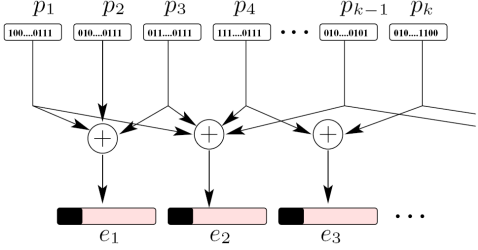

Fig. 1. Encoding at each source.

## IV. PROPOSED APPROACH FOR DATA DISSEMINATION IN VANETs

In this section, we introduce various aspects of our proposed scheme. First, we present the coding technique which play a key role in the dissemination problem. Next, we describe the packet transfer protocol, i.e., the set of rules that govern the packet transfer during contacts.

### A. Efficiency and Packet Transfer Technique

The broadcast data may be lost due to interference, packet collisions, and hidden node problems. The conventional approach to reduce collisions and the hidden node problem is using the request to send/clear to send (RTS/CTS) handshakes. This scheme can improve the data delivery ratio. However, bandwidth will still be wasted by back off timers, control messages, and RTS/CTS handshakes, which effectively reduce the dissemination throughput and increase the latency for time sensitive applications.

Vellambi et al. [14] showed that by applying rateless codes at the source where message is generated, even in the presence of packet expiry and intermittent connectivity, one can effect reliable message delivery with improved latency in the unicast scenario. For dissemination scenarios, likewise, we expect that we can gain similar improvements by applying rateless codes. To illustrate improvement in the overall throughput, we compare the following options.

Consider a data source that broadcasts a message consisting $N$ different data packets to a network of mobile nodes. Let assume that every node passing the source receives a portion of $N$ data packets. These nodes carry the packets as they move away from the source. Now, consider that a new node (referred as the collector) enters the network and moves toward the source. This node randomly picks up a number of message packets from carrier nodes as they meet. We are interested in determining the expected number of contacts (i.e., meeting with carrier nodes) that the new node needs to ensure that it has all the $N$ data packets. Let $T$ be the time (each contact is one epoch) needed by the collector to collect all $N$ packets. Let $t_i$ be the time elapsed to collect the $i^{th}$ packet after collecting $i-1$ packets. We note that the probability of collecting a new packet given $i-1$ packets have already been collected is $p_i = (N-i+1)/N$. Therefore, $t_i$ has geometric distribution with average $1/p_i$. Adopting the *coupon collector's problem* solution and using the approach suggested in [11], the expected number of contacts required by a vehicle approaching toward the source, such that it can collect all the required packets can be determined. We will use this in the following derivations.

**Uncoded Packetized Scheme:** In this scheme, the source broadcasts uncoded packets from the set of $N$ information packets repeatedly, in a round robin fashion. Therefore, carriers have random subsets of the $N$ packets. Then, we can write

$$E[T] = \sum_{i=1}^{N} \frac{1}{p_i} = 1 + \frac{N}{N-1} + \ldots + \frac{N}{1}$$
$$= N \sum_{k=1}^{N} \frac{1}{k} \approx N \ln N + \gamma N \quad \text{where } \gamma \text{ is a constant.}$$
(1)

Hence, in this scheme, $E[T]$ grows with $N \ln N$.

**Erasure-coding-based Scheme:** In this scheme, source first applies erasure coding on $N$ data packets and forms a set of encoded packets of size $N(1+r)$, where $r$ is the redundancy factor. Then, source sends the encoded packets in a round robin fashion. The original data can be recovered by collecting any subset of size $N$ from $N(1+r)$ coded packets. Then, the expected time needed for collecting $N$ packets is given by

$$E[T] = 1 + \frac{N(1+r)}{N(1+r)-1} + \ldots + \frac{N(1+r)}{N(1+r)-N}$$
$$= N(1+r) \sum_{k=Nr}^{N(1+r)} \frac{1}{k} \approx N(1+r) \ln(1+\frac{1}{r}). \quad (2)$$

In [15], authors considered schemes where data is first encoded with a replication factor of $r_p$ and then packetized into $sr_p$ chunks for some integer $s$. Simulations in [15] reveal the superiority of schemes using erasure-coding over simple replication. Although, there are several drawbacks with their approach. First, employing a fixed-rate erasure coding scheme raises this question as, "what rate is optimal?". As it is clear from (1) and (2), the higher the parameter $r$, the better the expected delay. The rational is that as $r$ gets larger, $E[T]$ gets closer to $N$. However, this comes with the cost of using more network resources such as bandwidth and storage capacity. Moreover, realistic assumptions such as finite packet expiry and time-varying channel losses make this scheme practically inefficient since they would require the scheme to be rate-adaptable.

### B. Infocast

In order to resolve the issues we discussed before, we introduce *DMRC*. In this approach, when a message arrives at an RSU, the RSU packetizes the message into smaller data packets. These packets are then encoded into a set of slightly bigger size using the described rateless encoding scheme. Then the RSU broadcasts the set of encoded packets. As shown in Figure 2, for each RSU, we divide the vehicles on the road into two classes:

**Collectors.** These are the vehicles that are approaching toward a specific RSU.

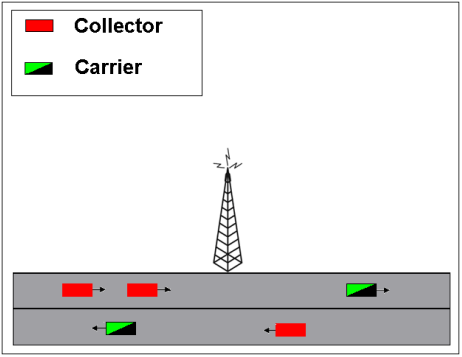

Fig. 2. For each source, we classify the vehicles on the road into two classes: *collectors* and *carriers*.

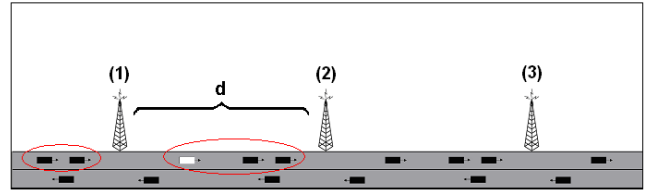

Fig. 3. Each carrier node can potentially carry packets from several sources simultaneously. It can act as a virtual source for an RSU and collector for other RSUs simultaneously. For example, the vehicle marked in white is carrier for (1) and collector for (2) and (3). The vehicles grouped in red form a cluster. Inter-cluster spacing is greater than communication range.

**Carriers.** These are the vehicles that have been successful in decoding a specific RSU's message (by collecting sufficient packets either directly from the RSU or from other carrier nodes). Note that as shown in Figure 2, the distance between carrier vehicles and the RSU increases with time. The carrier node broadcasts the RSU's message after encoding with rateless code as RSU. This would help collector nodes to recover the RSU's message much faster.

Collector vehicles switch to become *carrier* after they pass an RSU. Note that associated with every RSU, we attribute each vehicle as collector or carrier. Thus, it is possible that a single vehicle acts as a *carrier* for a specific RSU while the same vehicle acts as a *collector* for another RSU. A vehicle knows its location through GPS device and is pre-loaded with digital maps. Figure 3 shows the basic network model that we focus on; with a sequence of sources $\Phi_i$ placed uniformly with distance $d$ from each other. We refer to the space between two adjacent sources as *segments*. We say a node is in segment $\phi_{i,j}$ if the node is located in $j^{th}$ segment from source $\Phi_i$. Also, we denote $|\phi_{i,j}| = j$. Furthermore, we assume that all sources have equal amount of information packets of size $\mathcal{I}$.

In our dissemination problem, every node is interested in data from all the RSUs in the network. Each carrier node is potentially a new virtual source in the network and hence can encode the information packets and broadcast them to the collector nodes. In this scheme, each carrier node can potentially carry packets from several RSUs simultaneously. Thus, it can act as a virtual source and destination for different RSUs at the same time. Every time a collector node listens to a carrier node it receives packets which are innovative(by the rateless encoding property). That is, the encoded packets collected from various carriers are innovative although they are all coded from the same information packets belonging to an RSU. This is the best possible usage of contact opportunity which is an important factor in sparse scenarios. It can easily be shown that the expected number of required contacts is quite close to $N$. Hence, theoretically no extra overhead is needed by this scheme.

By using other specifications of VANETs, we can further improve the Infocast. We stress the location specific nature of the information (e.g., local video/audio news, store advertisements, as well as information about road conditions and car accidents). Hence, the information provided by RSU is only useful in a nearby geographical area and can be discarded outside that area. Our scheme is apt for incorporating this feature because carrier nodes can delete packets from far sources without any need for coordination with other nodes while maintaining the performance intact for nearby sources. To capture the spatial dependence of broadcast information, we define the relevance function for each source (RSU) as

$$\xi_j(\Phi_i) = (\Delta - |\phi_{i,j}|) \times d \qquad (3)$$

where $\Delta \times d$ is the maximum distance from an RSU that we want packets pertaining to an RSU $i$ present (i.e., *domain* of the RSU).

In the sequel we investigate the effect of two other important issues that play key roles in Infocast. First, we investigate the effect of vehicle mobility. By using the empirical data provided by [16], we study the effect of node clustering on the network parameters. We adapt the suggestion by [17] to employ multiple vehicles in parallel (when available) to deliver a relatively large file using the carry-and-forward approach. We then give an estimate of the distance from an RSU where a message can be recovered. This distance denoted by "Decoding Distance" (DD) is the most important metric in the performance evaluation of the Infocast.

### C. Impact of Mobility

One of the distinguishing features of vehicular networks is mobility. Disconnectivity is possible in such networks.Thus, clustering effect of nodes in the network must be investigated. Here, we first consider sparse highway scenarios where network connectivity is low. Let $S$ be the spacing between vehicles in a cluster. Empirical studies show that the inter-vehicle spacing may be assumed exponential with parameter $\lambda_s$ as [16]

$$f_S(s) = \lambda_s e^{-\lambda_s s}.$$

From this assumption it follows that vehicles form disjoint clusters. Vehicles in different clusters are far enough that they cannot communicate with each other.

**Theorem 1.** Let $V_0$ be the average speed of every vehicle on the road. Let $M_n(L)$ denote the number of clusters a collector vehicle meets during a travel along a road of length $L$ (note that collector and carrier vehicles move in opposite directions, with respect to an RSU). Further, let $M_t$ denote the time duration that a collector vehicle spends in contact with a cluster of carrier vehicles. Given that the inter-vehicle spacing follows an exponential distribution, we have

$$E[M_n(L)] \approx \frac{2L}{(e^{\lambda_s R}-1)(\frac{1}{\lambda_s} - \frac{Re^{-\lambda_s R}}{1-e^{-\lambda_s R}}) + R + \frac{1}{\lambda_s}} \quad (4)$$

and

$$E[M_t] = \frac{(e^{\lambda_s R}-1)(\frac{1}{\lambda_s} - \frac{Re^{-\lambda_s R}}{1-e^{-\lambda_s R}})}{2V_0} \quad (5)$$

*Proof:* We define $C_{length}$ and $S_{inter}$ as the cluster length and inter-cluster spacing. Then, it is easy to see that

$$E[M_n(L)] \approx 2L/E[C_{length}] + E[S_{inter}]$$

and

$$E[M_t] = E[C_{length}]/2V_0$$

The proof follows by using the statistical properties of clusters in VANETs [16] and results summerized in the following lemmas. We only give the proof for Lemma 4. The proof for the rest is straightforward. ∎

**Lemma 1.** Let $P_{last}$ be the probability that the distance between two successive vehicles is longer than the communication range $R$ (i.e., the probability of being the last vehicle in cluster). Then it can be shown that

$$P_{last} = Pr\{S > R\}$$
$$= 1 - F_S(R) = e^{-\lambda_s R}$$

**Lemma 2.** The PDF of inter-cluster spacing $S_{inter}$ (i.e., the space between the last vehicle of the leading cluster and the first vehicle of the following cluster) can be expressed as

$$f_{S_{inter}}(s_{inter}) = \lambda_s e^{-\lambda_s(s_{inter}-R)}$$

Therefore, we have

$$E[S_{inter}] = R + \frac{1}{\lambda_s}$$

**Lemma 3.** The expected number of vehicles in a cluster is given by

$$E[C_{size}] = \frac{1}{P_{last}}$$

**Lemma 4.** The average cluster length can be obtained as

$$E[C_{length}] = (e^{\lambda_s R}-1)(\frac{1}{\lambda_s} - \frac{Re^{-\lambda_s R}}{1-e^{-\lambda_s R}}) \quad (6)$$

*Proof:* One can show

$$C_{length} = \sum_{i=1}^{C_{Size}-1} S_i$$

Since $S$ follows an exponential distribution, we have

$$E[C_{length}] = E[E[C_{length}|C_{size}]]$$
$$= E[E[\sum_{i=1}^{C_{Size}-1} S_i]]$$
$$= E[C_{size} - 1]E[S|S < R] \quad (7)$$

Substituting the required expressions in (7) completes the proof. ∎

Results in Theorem 1 are consistent with the notion that as mobility increases, the number of encounters remains the same but the duration of encounters decreases. From (4) and (5), it can be shown that the expected number of packets that a collector vehicle obtains from $\Phi_i$ during travel in a segment $\phi_{i,j}$ is directly proportional to the number of packets from the corresponding source that carriers posses in that segment.

### D. Decoding Distance (DD)

The parameter $DD$ is basic performance metric we consider which can provide insight to the throughput. As mentioned earlier, each node has a limited buffer space for cooperation, i.e., each carrier node keeps only a limited number of encoded packets from sources it has met. Here, we address a simplified version of the problem of allocating the limited buffer to a number of sources. The objective is to maximize $DD$ for all sources.

Upon crossing an RSU, every node that has been successful in decoding the RSU's message act as a carrier for that source. Then, every carrier node generates some encoded packets from the RSU's information packets and stores them. The number of stored packets to be determined for maximum performance, given that the storage buffer is limited to $B$.

By meeting each cluster $C_i$, the collector has the opportunity to gather packets. Consider a collector vehicle meets a cluster of vehicles in $\phi_{i,j}$, the $j^{th}$ segment from $\Phi_i$. Assume that, on the average, there are $m_{i,j}$ packets from $\Phi_i$ in the buffers of the carriers in the cluster. Then, the number of broadcast packets during the meet time $M_t$, follows the Poisson distribution with mean $\rho$, because carriers send encoded packets randomly and without coordination. Thus, using the throughput relations of the ALOHA network [18], we verify that the maximum throughput occurs when $\rho$ is equal to $1/2$ of packet transfer time and is equal to $1/2e$. Further, the probability that a received packet is of collector's interest (i.e., the packet is from $\Phi_i$) is equal to $m_{i,j}/B$. Hence, by using (4) and (5), the maximum expected number of packets collected from a cluster $N_j^c$, when carriers posses $m_{i,j}$ packets from $\Phi_i$ and total number of collected packets $N_j^T$, that a typical collector can obtain from segment $\phi_{i,j}$ of length $d$, are given by

$$E[N_j^c] = \frac{(e^{\lambda_s R}-1)(\frac{1}{\lambda_s} - \frac{Re^{-\lambda_s R}}{1-e^{-\lambda_s R}})m_{i,j}}{4eV_0 B} \quad (8)$$

and

$$E[N_j^T] = M_n(d) \times E[N_j^c] \quad (9)$$

Therefore, $DD$ is directly proportional to the number of packets from the corresponding source that carriers posses per each segment. In order to maximize $DD$, we need to find a solution for $m_{i,j}$'s subject to the buffer limit constraint. Since sources are all the same, we can omit the first index in $m_{i,j}$ and find a generic solution $m_j$ for all sources. We assume that the number of packets a carrier posses from $\Phi_i$ cannot be increased. Also, buffer updating for a carrier node occurs when it crosses a new source and enters a new segment (e.g., from $\phi_{i,j}$ to $\phi_{i,j+1}$). Just after crossing $\Phi_i$ the carrier node has $m_0$ encoded packets from the source and reduces them gradually as $m_0 \geq m_1 \geq m_2 \geq \cdots$ and $m_j = 0$ for $j \geq \Delta$. To find the maximizer distribution for $m_j$'s, we use the symmetricity in the problem and consider a collector which enters the source $\Phi_i$'s domain. Hence, $DD$ can be formally stated as

$$DD = \min_{m_j} d \quad (10)$$

$$s.t. \quad \sum_{i=0}^{d} E[N_{\Delta-i}] \geq \mathcal{I},$$

where $E[N_{\Delta-i}]$ is the expected number of packets obtained by a collector in segment $(\Delta - i)$. In other words, we are looking for the first segment that the total number of collected packets is greater than $\mathcal{I}$. Since (10) suggests that only the tail behaviour of the distribution of $m_j$'s is important and because $m_j$ is non-increasing with $j$, one can see that the maximum value of $DD$ is achieved when $m_\Delta$ (and hence all the previous segments) has its maximum value. Further, the buffer limit constraint implies that $\sum_{i=0}^{\Delta} m_i \leq B$. Therefore, a solution can be formulated as

$$m_0 = m_1 = m_2 = \cdots = m_\Delta = \frac{B}{\Delta + 1} \quad (11)$$

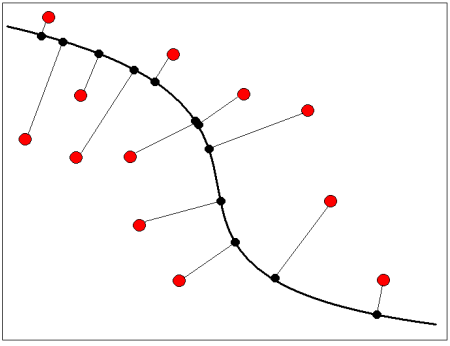

Fig. 4. Two-Dimensional Scenario

| Parameter | Value |
|---|---|
| Simulation time | 1000 seconds |
| Communication range (R) | 200 m |
| Vehicle velocity | 20–40 m/s |
| Inter-arrival of vehicles ($\lambda$) | 0.1 veh/second |
| Simulation road length | 20000 m |
| Number of nodes | 200 |
| Broadcast interval | 100/second |
| Drop factor (**D**) for scheme A | 5–30 |
| Window size (**N**) for scheme B | 1–50 |
| Buffer size | 100–1500 |
| Number of RSU's along the road | 50 |

TABLE I
SIMULATION SETUP

The result on a single linear highway can be extended to the two-dimensional case, using the trajectory of vehicles, we can still classify nodes into two different groups as collector nodes and carrier nodes. Then, as depicted in Figure 4, we can transform a two-dimensional scenario into a one-dimensional case by mapping sources on nodes' path. Therefore, the analysis is quite similar to the one-dimensional case but with non-uniformly distributed sources.

## V. RESULTS OF SIMULATIONS AND DISCUSSIONS

### A. Simulation setup

In this section, we evaluate the performance of a simple Infocast setup. Vehicles enter the road from one end with inter arrival times drawn using instances of exponential distribution with parameter $\lambda$. We considered a random placement of 50 points on a 2–D plane which are mapped on the diameter as the positions of the RSUs along the road. Using the results from IV-D we suggest two different buffer management schemes and evaluate their performance metrics.

We developed an NS-2 [19] based simulator to evaluate the proposed schemes. Vehicles can communicate with each other via short-range communication transmission. We use the transmission range $R$ as the unit of distance in our evaluations. We are interested in collector vehicles who move toward the source. We focus on a collector vehicle departs from one end of the road and travels along it until it reaches the other end. The record of all the collected packets versus distance is maintained for analysis. We assume that each source has $\mathcal{I}$ innovative information packets and broadcasts rateless encoded packets repeatedly. Further, each carrier is assumed a buffer of size $B$. Most experiment parameters are listed in Table I.

The performance of the protocols is measured by the following metrics:

- *Mean Decoding Distance (MDD),* which is the expected value $DD$ over all active sources
- $P_{Success}$ is the probability that a random message generated at a (random) source $\Phi$ is available at node $v$ before it enters the communication range of $\Phi$. We measure $P_{Success}$ as a function of distance to the source. The presented graphs are the average of $P_{Success}$ over all active sources
- *Deployment Capacity (DC),* which is the maximum number of active sources on the road such that the $P_{Success}$ for a collector vehicle at an average distance $\eta$ from a source is in $\epsilon$ neighborhood of 1 ($\epsilon \ll 1$).

In order to meet the optimal distribution of $m_j$'s as presented in (11) and compare the effect of different parameters,

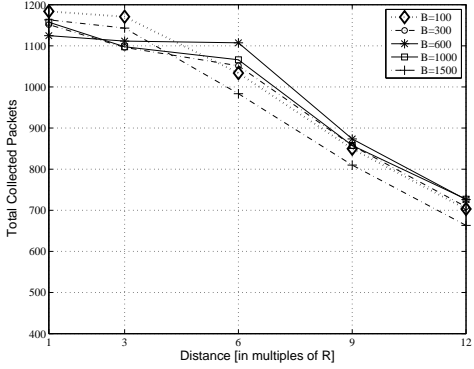

Fig. 7. Total number of collected packets vs distance from an arbitrary source for different values of the buffer size in scheme B

|  | $\eta = R$ | $3R$ | $6R$ | $9R$ | $12R$ |
|---|---|---|---|---|---|
| M=10 | 1817.2 | 1699.9 | 1570.8 | 1462.7 | 1395.3 |
| M=20 | 915.9 | 832.65 | 754.6 | 677.55 | 628.3 |
| M=30 | 708.5 | 623.73 | 568.63 | 452.37 | 400.63 |
| M=40 | 754.7 | 677.48 | 590.1 | 446.33 | 388.63 |
| M=50 | 933.2 | 662.31 | 608.39 | 472.95 | 423.68 |

TABLE II
DEPLOYMENT CAPACITY EVALUATION–SCHEME A

|  | $\eta = R$ | $3R$ | $6R$ | $9R$ | $12R$ |
|---|---|---|---|---|---|
| M=10 | 2068 | 2067.4 | 1997.4 | 1825.1 | 1673.7 |
| M=20 | 1459.2 | 1450.5 | 1267 | 1080.2 | 937.05 |
| M=30 | 1196.4 | 1152.3 | 1128.3 | 915.9 | 779.8 |
| M=40 | 1117.6 | 1059.3 | 1031.7 | 831.8 | 709.1 |
| M=50 | 1091.1 | 1134.6 | 1007.3 | 810.3 | 700.85 |

TABLE III
DEPLOYMENT CAPACITY EVALUATION–SCHEME B

two different buffer management schemes are proposed for simulations.

*1) Scheme A:* In this scheme, the buffer space is shared by *all the sources* (RSUs) a vehicle has met thus far. When reaching a new source, a fraction **D** of the packets in the buffer are dropped to store the new source packets. Upon crossing the new source and entering a new segment, fraction **D** of the packets from every source will be dropped randomly.

*2) Scheme B:* In this scheme, the buffer space is shared by a fixed number of *the most recent sources* a carrier has met. Upon reaching a new source, all the packets from the oldest source in the buffer will be removed and replaced by the incoming packets from the new source.

We perform the simulation for different values of **N**, the number of sources from which a carrier keeps packets in its buffer (**N** = 1 is equivalent to flushing the buffer upon reaching every new source). Figure 5 compares the MDD for both schemes. As Figure 5(b) suggests, for high values of **N**, scheme B outperforms scheme A. This result conforms the intuitive solution presented in (11). Because, for large values of **N**, the carrier node keeps packets form previous sources for a longer amount of time and in fact shares the available buffer space on the road between all sources in a window of size **N**. When **N** is large, packets from far source stay longer at the carrier buffer and hence giving more chance to the collectors to gather packets from the far source. It is worth noting that large **N** results in relative fairness between sources.

To plot $P_{Success}$, we set the value $\mathcal{I}$ first and count the number of collected packets $P_\eta$ (versus distance $\eta$ from the source) that a collector node can accumulate before entering the communication range of the source. Define the indicator random variable as

$$\mathcal{S}_\eta = \begin{cases} 1 & \text{if } P_\eta \geq \mathcal{I} \\ 0 & \text{if } P_\eta < \mathcal{I} \end{cases}$$

The value of $Pr\{\mathcal{S}_R\}$ is considered as $P_{Success}$. Figure 6 shows $P_{Success}$ for a number of distances (in multiples of communication range R) from an arbitrary source. The collector nodes would like to recover the source information from a far distance. One observation to make from Figure 6(b) is that in scheme B, by using a larger value of **N** we can ensure the success for collector nodes. Hence, scheme B provides a very good coverage area around each source by enabling collector nodes to decode the source message with high probability even before entering its communication range.

In section IV-D it was claimed that the expected value of collected packets has linear dependency on $\frac{m_j}{B}$. In order to verify that, in Figure 7, we have plotted total number of collected packets for a collector vehicle for different values of $B$ in scheme B. Since in scheme B the buffer space is equally divided among all the **N** sources within the window, the fraction $\frac{m}{B}$ remains the same. Hence, although there are large variations in buffer size, in Figure 7, we only see subtle changes in the total number of collected packets and the trend remains the same for all of them.

Tables II, III show the average of total number of collected packets per source for a collector node as a function of distance $\eta$ and different number of active sources ($M$) on the road. To determine the "Deployment Capacity" (DC), we need to fix a distance and a positive value $\epsilon$ which indicates the bound on acceptable $P_{Success}$. Then, by using the tables we are able to find the range of the maximum number of sources that satisfies the condition on decoding probability at desired distance. Then, $DC$ can be obtained by performing more simulations on that range and altering the number of active sources incrementally. Suppose that we are interested in $DC$ for scheme B at $\eta = 6R$ with $\epsilon \ll 1$ when $\mathcal{I} = 1000$. Since total number of collected packets per source when $M = 50$ is slightly higher than $\mathcal{I}$ (from Table III), we conclude that $40 < DC(\epsilon) < 50$. Using the graph in Figure 8, we conclude that $DC(\epsilon < 0.05) = 44$, $DC(\epsilon = 0.05) = 46$ and $DC(\epsilon = 0.25) = 48$.

## VI. CONCLUSIONS AND FUTURE WORK

We presented a novel scheme based on rateless codes for collaborative content distribution from road side units to vehicular networks. In Infocast, information sources in the network broadcast the encoded data repeatedly. In Infocast, we divide the vehicles on the road into two groups of collector and carrier nodes per source. The collector nodes have the opportunity to recover the source message from a far distance using the carrier nodes. We introduced Decoding Distance and Deployment Capacity for performance evaluation. We also provided analytical models to explore the Decoding Distance and Deployment Capacity of the proposed dissemination schemes. The presented models capture the effect of various parameters in the network and provide guidelines on choosing the parameters to maximize the performance metrics.

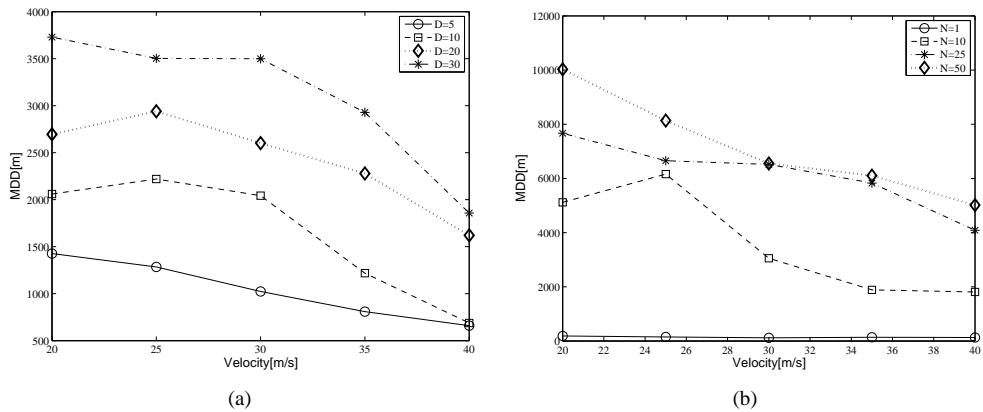

Fig. 5. Mean Decoding Distance (MDD) vs. velocity for different values of **N** and **D**: (a) scheme A, (b) scheme B

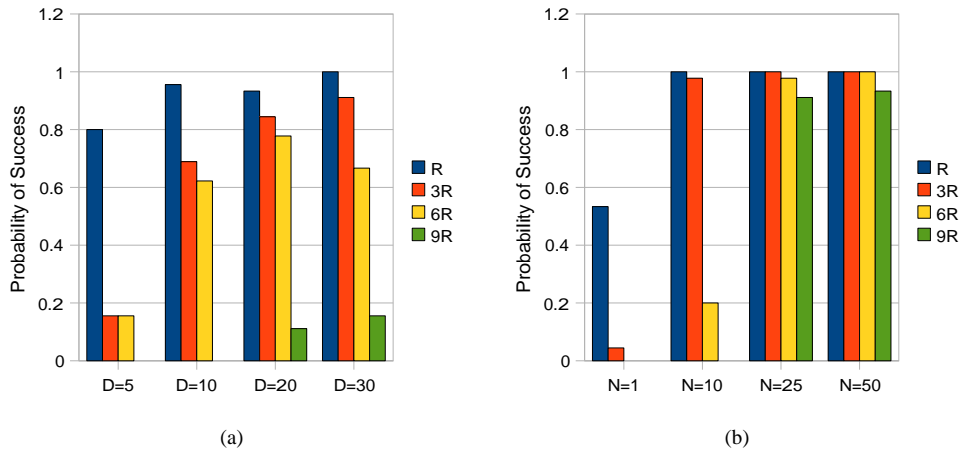

Fig. 6. Probability of Success for various distances (in multiples of the communication range, R) from source: (a) scheme A, (b) scheme B

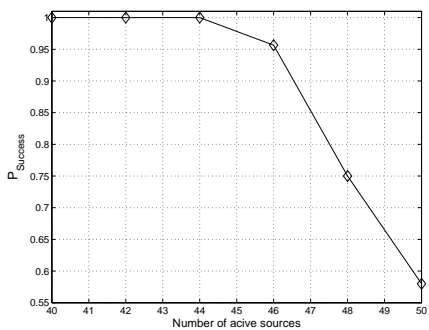

Fig. 8. $P_{Success}$ vs number of active sources in scheme B for $\eta = 6R$

To the best of our knowledge, this is the first paper to study the effect of finite buffer constraint in the dissemination problem in a VANET. Infocast proposed as a general solution for improving the reliability and Decoding Distance. The proposed scheme can seamlessly handle both sparse and dense scenarios. Future directions are to introduce more realistic traffic models for urban areas where the exponential assumptions for inter-arrival time is not valid. Further, adapting the analysis for such scenarios and finding the optimal distribution for buffer allocation are our immediate goals. Finally, deriving better approximations for performance metrics in more general setting and relaxing some of the assumptions remains as open problems.